\documentclass{JHEP5}
\usepackage{amsfonts,amsbsy,latexsym,amssymb,amscd,amstext}
\usepackage{epsfig}
\usepackage{graphicx}
\newcommand{\be}{\begin{equation}}
\newcommand{\ee}{\end{equation}}
\newcommand{\ba}{\begin{array}}
\newcommand{\ea}{\end{array}}
\newcommand{\baa}{\begin{array}}
\newcommand{\eaa}{\end{array}}
\newcommand{\bea}{\begin{eqnarray}}
\newcommand{\eea}{\end{eqnarray}}

\newcommand{\bsigma}{\overline{\sigma}}
\newcommand{\Dslash}{\not \! \! D}
\newcommand{\RF}{R^4}
\newcommand{\OMF}{\omega}
\newcommand{\omp}{\delta_1}

\title{Gluino zero-modes for non-trivial holonomy calorons}

\author{Margarita Garc\'{\i}a P\'erez $^{a}$ and Antonio Gonz\'alez-Arroyo $^{a,b}$
 \\
  $^a$ Instituto de F\'{\i}sica Te\'orica UAM/CSIC,  C-XVI \\
  $^b$ Departamento de F\'{\i}sica Te\'orica, C-XI \\
       Universidad Aut\'onoma de Madrid, E-28049--Madrid, Spain \\

E-mail: \email{margarita.garcia@uam.es, antonio.gonzalez-arroyo@uam.es
  }}

\abstract{
We couple  fermion fields in the adjoint representation (gluinos)  to the 
SU(2) gauge field of  unit charge  calorons defined on $R^3\times S_1$.
We compute corresponding  zero-modes of the Dirac equation. These are 
relevant in semiclassical studies of ${\cal N}=1$ Super-symmetric Yang-Mills
theory. Our formulas, show that, up to a term proportional to the vector
potential, the modes can be constructed by   different linear
combinations of two contributions adding up to the total caloron field
strength.  
}

\preprint{IFT-UAM/CSIC-06-27\\ FTUAM-2006-15}

\begin{document}

{\vskip 1cm}

\section{Introduction}
\label{s.intro}

In recent years a deep link between monopoles and instantons has
been established at finite temperature \cite{kvanbaal1}-\cite{falk}. It was 
for long 
known that the finite temperature instanton, the Harrington-Shepard caloron 
\cite{HS}, becomes for large scale parameter a BPS monopole 
\cite{Rossi}. This happens when the value of the constant Polyakov loop at 
infinity, the  holonomy, approaches unity. The surprise came with the 
discovery that non-trivial holonomy calorons when squeezed in the
time direction reveal to be composed of $N$, for $SU(N)$, 
BPS monopoles. $N-1$ of these constituent monopoles 
are massless for the HS caloron, but in general all of them get non-zero masses
with values related to the eigenvalues of the Polyakov loop at infinity.
The idea of the composite nature of instantons, with  
instanton quarks \cite{quarks} or  merons \cite{merons} as constituents, 
has been on the basis of several  semiclassical 
proposals to address the confinement problem in QCD. Isolated
fractional instantons (twisted instantons)\cite{usfrac} can be obtained 
by using tori with twisted boundary conditions \cite{thooft}. By
replicating the tori one can obtain classical configurations in a
periodic box, where the action density is clustered into lumps of 
$1/N$ of topological charge. These structures were observed in lattice
generated ensembles at zero temperature and were argued to be relevant for QCD  
confinement \cite{tonyfrac}. 
Non-trivial holonomy calorons also exhibit explicitly this composite nature 
as far as the separation between constituents stays larger that their size.
Otherwise they merge in an undissociated instanton. 
Triggered by this result,  quite a number of more recent lattice analysis  
have identified the  presence of constituent monopoles at temperatures below
but close to the deconfinement phase transition in Monte-Carlo generated configurations
\cite{gattringer1,lattice}.
The possible relevance of the instanton-monopole link for QCD dynamics is an 
open issue, nevertheless constituent monopoles have already shown their 
usefulness in a different context, in particular for calculations of the gluino 
condensate in 4D ${\cal N}=1$ supersymmetric Yang-Mills theory \cite{hollowood}.
The contribution of the constituent monopoles seems essential there to bring to 
agreement strong and weak-coupling instanton calculations of the 
condensate.

The existence of an analytic expression for the non-trivial holonomy calorons
allowed a subsequent analytic calculation of the zero-modes of the Dirac 
equation, in the fundamental representation, 
in this background field\cite{us,maxim}. For large constituent separation
the  modes are  entirely supported on just one of the monopoles, jumping from one
to other as we change the periodicity condition in the time-like direction imposed on 
the solution. This knowledge has proven useful in interpreting the
results of several lattice studies which employ  low-lying eigenstates of the 
Dirac operator  to trace topological structures present in gauge field 
configuration ensembles \cite{gattringer1}, \cite{gattringer2}-\cite{falkpierre}.

The present paper is devoted to the derivation of the analytic expression 
and properties of the zero-modes of the  Dirac equation in 
the adjoint representation for Q=1 SU(2) calorons. These are 
relevant objects in the study 
of the semiclassical behaviour of  4D ${\cal N}=1$ supersymmetric
Yang-Mills theory compactified in $R^3 \times S^1$. They are directly 
related to estimates of the gluino-condensate~\cite{cohen-gomez}. 
By now this has been studied~\cite{hollowood} only
in the limit of large constituent separation, in which  the SU(2) caloron
degenerates into two BPS monopoles. Our expressions for the
four adjoint modes, valid for any separation, indeed tend in this limit 
to two pairs, one  attached to each constituent monopole reproducing  
the well known adjoint zero modes of BPS monopoles \cite{Rossi}.

Our work is also useful within the  previously mentioned spirit 
of using modes  of the Dirac equation as probes of gauge field structure,
an approach that has become very popular (see for instance \cite{diraclatt00}-
\cite{diraclatt7})
since the discovery of lattice Dirac operators \cite{neuberger} that
possess
exact index theorems.
The usefulness of  adjoint modes in this respect has  been recently 
advocated~\cite{tonyadj}.
The idea is based upon the so called supersymmetric modes, having densities 
that match the action density profile but are less sensitive to ultraviolet
fluctuations.

The paper is organised as follows. In section 2 we describe the general 
properties of adjoint zero-modes, such as their occurrence in pairs, related 
by Euclidean CP transformations. Then, we give  the main formula for the 
modes  within the ADHM formulation, and  apply it to the $Q=1$ SU(2) 
caloron case. This provides two pairs of zero-modes. One pair is given by 
the supersymmetric zero modes, which are proportional to the gauge field
strength itself. The remaining  pair of adjoint zero modes is studied. 
In addition to the analytical expression (details of its derivation are
given in the Appendix), we display  its density profile 
in some representative cases. In section 3 we show how the solutions behave 
in certain limits and how they interpolate between modes of BPS monopoles and 
those of instantons.
  We end up with a summary of the  results and a list of  possible
  extensions and applications.  

\section{Formalism}
\label{s.form}

The adjoint zero modes $\Psi^a_\alpha(x)$ are solutions of the Euclidean 4-dimensional 
massless covariant Dirac equation in the adjoint representation of the gauge group:
\be
\Dslash \Psi=0
\ee
In this paper we will analyse the simplest case given by group SU(2). Then
the colour index $a$ takes the values $1,2,3$, while the spinorial index 
$\alpha$ takes four values. Action by $\gamma_5$ maps zero-modes into other
ones. It is convenient then to combine  zero modes into  eigenstates of 
$\gamma_5$ with eigenvalue $ \pm 1$, known as left and right-handed modes
respectively. In Weyl's representation of the Dirac matrices, the left and
right handed modes  reduce to two-component spinors $\psi_{\pm}$ satisfying 
\bea
\hat{D} \psi_- & =& 0 \label{rhequation}\\
\bar{D} \psi_+ & =& 0 \label{lhequation}
\eea
where $\hat{D}= \sigma_\mu D_\mu$ and $\bar{D}= \bsigma_\mu D_\mu$.
The Weyl matrices are given by $\sigma_\mu= (\mathbf{I}, -i\vec{\tau})$, 
in terms of  the Pauli  matrices  $\tau_i$ (the matrix $ \bsigma_\mu$
is the hermitian conjugate of $\sigma_\mu$). If the gauge field is self-dual,
then Eq.~\ref{rhequation} implies 
\be
D_\mu \psi_- = 0 
\ee
for all $\mu$. This can easily be shown to imply that the gauge-invariant
density $|\psi_-(x)|^2$ must be constant (x-independent). For non-compact 
space-times these are non-normalizable solutions.

Focusing now on left-handed zero modes (solutions of Eq.~\ref{lhequation}) 
we point out  that the space of solutions is always even-dimensional. This
follows from euclidean CP invariance mapping one solution into  other 
\be
\psi_+ \longrightarrow \psi_+^c  \equiv -i  \tau_2 \psi_+^\dagger
\ee
In the previous formula $\psi_+^\dagger$ stands for the complex conjugate
spinor and the matrix $\tau_2$ acts on the 2-spinor indices. Furthermore, 
for self-dual gauge fields one can establish a one to one correspondence
between self-dual deformations $\delta A^a_\mu$ of the gauge field and 
left-handed zero  modes. Given a deformation, one must first transform it 
to the background Lorentz gauge
\be
D_\mu \delta A_\mu=0 
\ee
and then one can show that 
\be
\psi_+= \delta A_\mu \sigma_\mu V
\ee
is a zero mode for any constant 2-spinor $V$.

If there are  isommetries of the problem which do not leave the solution
invariant, the corresponding deformations are associated to specific
zero-modes. In particular, the  {\em super-symmetric zero-modes} arising  
for $\delta A_\mu= F_{\mu 0}$,  are associated with  translation 
symmetry. In general, the space of self-dual connections is continuous 
and can be parameterised in terms of a set of real parameters (moduli). 
Variations with respect to these moduli parameters(tangent vectors) 
give rise to adjoint zero-modes. 

Self-dual gauge fields with topological charge $Q$  on the sphere
$S_4$ or on  $\RF$ (with finite action)
can be constructed by an algebraic procedure known as the ADHM
construction~\cite{adhm}. The fields are written  in terms of  $q$, 
a Q dimensional column vector of quaternions, and $\widetilde{A}$, a
$Q\times Q$ matrix of quaternions satisfying certain conditions, to be 
specified later.  In what follows, we will identify quaternions with 
the space of two by two matrices which are real linear combinations of the Weyl matrices
$\bsigma_\mu$. In particular, one can form the quaternion $\hat{x}=x_\mu
\sigma_\mu$ and its adjoint  $\bar{x}=x_\mu
\bsigma_\mu$. The self-duality condition amounts to the requirement  that the matrix $R$:
\be
R= q\otimes q^\dagger +
(\widetilde{A}^\dagger-\bar{x})(\widetilde{A}-\hat{x})
\ee
is real and invertible.

The self-dual deformations are then associated with variations of the ADHM
data $q$ and  $\widetilde{A}$ ~\cite{osborn}.  Using the previously mentioned 
relation between
adjoint modes and deformations, one obtains the formula for the modes 
in terms of variations of the parameters $\delta q$ and $\delta
\widetilde{A}$:
\be
\label{zero_mode}
\delta A_\mu= \frac{-i }{2} (\delta q^\dagger - u^\dagger
\delta \widetilde{A})
\bar{\sigma}^\mu \hat{\partial}\OMF + \mbox{h.c.}
\ee
where we have introduced the x-dependent vectors of quaternions $u$ and
$\OMF$ defined by the relations
\be 
\label{OMeq}
\OMF= R^{-1} q
\ee
and 
\be
\label{ueq}
u=F\,  (\widetilde{A}-\hat{x}) \OMF
\ee
where $F=1+u^\dagger u$ is a real  function.  The symbol $\hat{\partial}$ in
Eq.~\ref{zero_mode} stands for the contraction $\partial_\mu \sigma_\mu$. 

The condition that the variation $\delta A_\mu$ given in  Eq.~\ref{zero_mode}
satisfies the required covariant
background gauge condition is 
\be
\label{RBGC}
{\rm Re}\left( \widetilde{A}^\dagger\,  \delta \widetilde{A} - \delta
\widetilde{A}^\dagger\,   \widetilde{A} +q(\delta q)^\dagger -(\delta q)
q^\dagger \right)=0 
\ee
where ${\rm Re}$ stands for the real part of the quaternion.  This condition 
will be seen to hold for our formulas. Its interpretation will also become 
more clear later.

Now we will  proceed to particularise to  the case of the Q=1 caloron. 
This is a self-dual  configuration in $R^3\times S_1$. At infinity the 
time-like Polyakov loop (the holonomy) is non-trivial.
This is determined by the parameter $\omp$ such that the trace of the
Polyakov loop at spatial infinity tends to $\cos( 2 \pi \omp)$.

For a given holonomy and a fixed period in time $\beta$ (which we will
henceforth fix to 1), solutions depend on the following parameters: the position 
of the center of mass of the caloron $X_{\rm  CM}$,  the size parameter 
$\rho$ and an SU(2) colour orientation. For medium and large values of $\rho$
(compared with the time-period $\beta=1$) the action density
of the solution appears as a  superposition of two lumps, named 
{\em constituent monopoles} in Ref.~\cite{kvanbaal1}. 
The distance between the lumps approaches $\pi\rho^2$ and the total
masses $M_a= 4 \pi m_a/g^2$  are 
given by the holonomy as follows:
\be
m_1= 4\pi \delta_1 \quad ;\quad m_2= 4 \pi \delta_2=2 \pi - m_1 
\ee
The shapes of these monopoles tend, as the distance is
increased, to that of  BPS monopoles, having a non-abelian core  which is
exponentially localised and an abelian powerlike fall-off at large distances.

One can construct the caloron solution by an infinite dimensional
generalisation of the ADHM construction~\cite{kvanbaal1} (strictly speaking a
Nahm~\cite{nahm} transform). This is the procedure that we will follow here, 
allowing us to extend the ADHM formulas for the adjoint modes to this case. 
If we regard the caloron solution as a  solution in  $\RF$ which is periodic
in time, the topological charge would now become infinite. Thus, 
$q$ and $\widetilde{A}$ become an infinite dimensional vector and matrix 
respectively. The discrete index  can be interpreted as the Fourier mode
of a periodic function of one variable $z$ (with period 1). 
Thus $q(z)$ is a distribution and $\widetilde{A}$  a linear operator 
in this space. One can use 
translations, rotations and gauge transformations to bring $q(z)$
and $\widetilde{A}$ to the form ~\cite{kvanbaal1}:
\bea
q(z)=\rho (\delta(z-\delta_1)P_+ + \delta(z+\delta_1) P_-)\\
\widetilde{A}(z)=\frac{1}{2\pi i }\frac{d }{dz}  - i \vec{X}^1 \vec{\tau}
\chi_1(z) - i \vec{X}^2  \vec{\tau} \chi_2(z) 
\eea
where $P_\pm= \frac{1 \pm \tau_3}{2}$, $\delta(z)$ are  periodic 
delta functions and $\chi_a(z)$ are characteristic functions of the 
intervals $I_a$ (taking the value 1 in the interval and zero elsewhere). 
The intervals $I_1=[-\delta_1,\delta_1]$ and $I_2=[\delta_1,1-\delta_1]$ denote 
complementary regions of length $m_a/(m_1+m_2)$ within one period in $z$. 
Finally, the vectors $\vec{X}^a$ can be interpreted as  denoting the spatial
locations of the constituent monopoles. We have used the translation and 
rotation symmetry  to place them along the $z$ axis and to locate their
center of mass at  the origin ($m_1\vec{X}^1+ m_2\vec{X}^2=0$).
In addition, their separation 
is fixed by $\rho$:
\be
\vec{X}^2-\vec{X}^1= \pi \rho^2  (0,0,1)
\ee
This information allows to determine $\vec{X}^a$ uniquely. As mentioned all
$Q=1$ caloron solutions can be obtained from these formulas by applying 
euclidean and gauge transformations. Furthermore, one can easily 
restore of arbitrary time period  by multiplying all length 
parameters by $\beta$ (and masses by $1/\beta$). 

Within the Nahm transform philosophy, the quantity $\widetilde{A}$ can be 
identified with the covariant derivative (divided by $2 \pi i$) of an abelian 
gauge potential $\widehat{A}_\mu$ over a  4-d torus which has been shrunk 
to a circle, whose coordinate is labelled by $z$. The remaining (spatial) 
coordinates have dropped as arguments, but  the vector potential  field
still keeps the vector index.
In our case only the third component is non-zero, and is given by
\be
\widehat{A}_3= - 2 \pi (X^1_3 \chi_1(z) + X^2_3 \chi_2(z))
\ee
This implies that the corresponding magnetic field vanishes and the electric 
field is a delta function over $z=\pm \delta_1$.

In conclusion, to obtain the expression  for the adjoint zero modes for  our
case, one has only to substitute the expression of the variations 
$\delta q $ and $\delta \widetilde{A}$ in the formula Eq.~\ref{zero_mode}.
The variations are associated to the parameters of which the caloron field 
depends. On one hand we have the coordinates of the center of mass of the 
constituent monopoles. This will give rise to the supersymmetric modes, which 
are always associated to translational symmetry. The corresponding
variations can be obtained straightforwardly, giving $\delta^{(0)} q=0$ and 
$\delta^{(0)} \widetilde{A}= \mathbf{I}$. Substituting into the formula
one obtains $\delta^{(0)} A_0 =0$ and 
\be
\delta^{(0)} A_i = E_i
\ee
where $E_i$ are the components of the electric (or magnetic) field strengths 
of the caloron. As anticipated this  is the expression of the supersymmetric
mode, having  density proportional to the action density of the caloron field. 

The index theorem suggests that we should find 4 independent solutions ($2N Q$).
As mentioned previously they come in CP-pairs. Each pair is associated 
to 4 real variations of the Nahm data. This is exemplified with the
supersymmetric modes, for which there is a single pair associated to the 4d 
center of mass variations. Therefore, one must still find a new independent 
CP-pair. As we will  see,  one can obtain one such mode by varying with 
respect to the parameter $\rho$. Acting with the operator $\rho
\frac{d}{d \rho}$  on our expressions of  $q$ and $\widetilde{A}$, we 
obtain $\delta^{(1)} q= q$ and 
\be
\delta^{(1)} \widetilde{A}=  - 2i X_3^1 \tau_3
\chi_1(z) - 2i X_3^2  \tau_3 \chi_2(z)=  i \rho^2 \tau_3 (m_2 \chi_1(z) -m_1
\chi_2(z))
\ee
For these variations $q (\delta^{(1)} q)^\dagger - (\delta^{(1)} q) q^\dagger$
vanishes, and  Eq.~\ref{RBGC} amounts to the requirement  that the Nahm-dual 
field $\widehat{A}_\mu(z)$ satisfies the covariant background gauge condition too.
This follows trivially, since $\delta^{(1)} \widetilde{A}_0=0$. In summary, 
substitution of 
these variations into Eq.~\ref{zero_mode} provides a self-dual deformation 
in the covariant background Lorentz gauge.  The same applies to the
supersymmetric zero-mode for which $\delta^{(0)} q=0$   and $\delta^{(0)} \widetilde{A}_0$ is constant.  

To give a simple expression of the result it is convenient to separate the 
contribution of $\delta q$, present only in the non-supersymmetric zero-mode
case from the other one. Furthermore, we point out that the former becomes
proportional to the caloron vector potential itself:
\be
\frac{i}{2} q^\dagger \bar{\sigma}_\mu \hat{\partial}\OMF + \mbox{\rm h. c.} =
\frac{2}{F} A_\mu 
\ee

To give an expression of the second term one must realize that both $u$ and 
$\OMF$ can be written as the sum of two contributions, $u=\sum_a u_a$ and
$\OMF= \sum_a \OMF_a$, such that each piece is proportional to the 
characteristic function $\chi_a(z)$  of each of the intervals. 
Finally, we can collect the formula for both sets of modes as follows
\bea
\label{susy}
\delta^{(0)} A_\mu &=&  E^1_\mu +E^2_\mu  \\
\label{nonsusy}
\delta^{(1)}A_\mu &=& - \frac{2}{F} A_\mu + \rho^2 \eta^{3 \mu}_\alpha (m_2
E^1_\alpha -m_1 E^2_\alpha)
\eea
where $\eta^{3 \mu}_\alpha$ is `t Hooft symbol and we have defined 
\be
E^a_\alpha = \frac{i}{2} u_a^\dagger \bar{\sigma}_\alpha \hat{\partial}
\OMF_a + \mbox{\rm h. c.}
\ee
which by virtue of Eq.~\ref{susy} can be regarded as the contribution 
of constituent monopole {a} to the caloron field strength. 
This formula is quite appealing since it suggests that the two modes 
are simply given by  different linear combinations of 
the field produced by each constituent monopole. 
This is modified by the presence of the $\eta^{3 \mu}_\alpha$ in the 
expression for the non-supersymmetric mode. Notice, however, that 
each mode belongs to a complex two dimensional space of modes generated 
by the two elements of a CP-pair. In  particular, this means that one can 
transform the gauge variations as follows
\be
\delta' A_\beta= \eta^{\alpha \mu}_\beta \delta A_\mu 
\ee
for any value of $\alpha$. Using this and allowing for a different
normalisation 
one can recast the formula for the 
non-supersymmetric mode as follows:
\be
\delta^{' (1)}A_\mu = - \eta^{3 \alpha}_\mu \frac{1}{\pi F \rho^2}
A_\alpha +   (m_2 E^1_\mu -m_1 E^2_\mu)/{2 \pi}
\ee

In the appendix we compute the  expressions for the functions $u_a$,
$\OMF_a$ and with them we compute $E^a_\alpha$. 
The final expression Eq.~\ref{Eformula} is a sum  of two terms. The first 
one is given by the field produced by a BPS monopole, of mass $M_a$,
located at the position of the corresponding constituent monopole, 
gauge rotated and weighted by an x-dependent scalar function $\lambda_a(x)$. 
In the next section we will analyse how these functions behave in different 
limits and reduce to the formulas for monopoles and instantons. 

All of the expressions are dependent on a  $2 \times 2 $ matrix
which is additive with respect to the contributions of the intervening 
monopoles. Its inverse, labelled  $V$, appears in the formulas and provides 
the main effect of one constituent monopole over the other.

\FIGURE{
\centerline{
\psfig{file=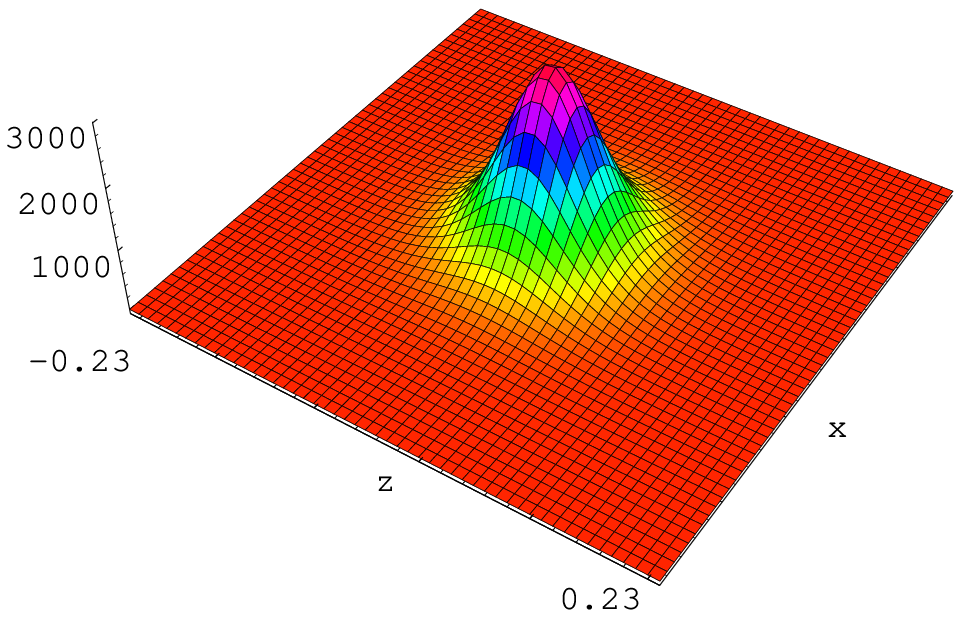,angle=0,width=8cm}
\psfig{file=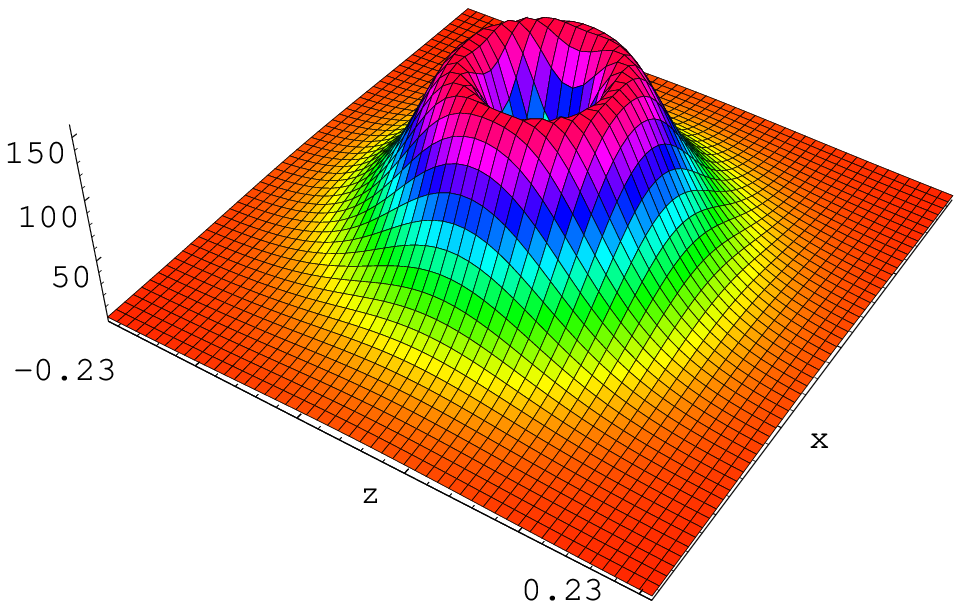,angle=0,width=8cm}}
\caption{Supersymmetric (left) and non-supersymmetric (right) zero-mode densities
for $\delta_1=0.2$ and $\rho=0.1$. Monopoles are localised on the
z axis  at $x=y=t=0$. Densities are plotted in the $x-z$
plane keeping $t=0$ and $y=0$. The lengths of the $x$ and $z$  axes are
scaled by $4.6\rho$.}
\label{fig1}}

As illustration of the properties of the zero modes we show in
Figs. (\ref{fig1})-(\ref{fig3}) the densities of both supersymmetric and
non supersymmetric zero modes for $\delta_1=0.2$ and several values of the 
scale parameter $\rho$. Densities are plotted in the $x-z$ 
plane keeping $t=0$ and $y=0$. 
For small $\rho$ the supersymmetric zero-mode reproduces the characteristic 
single-instanton shape. As $\rho$ increases the caloron dissociates
into two constituent monopoles which tend, at large $\rho$, 
to two BPS monopoles. The non-supersymmetric mode 
has, at small $\rho$,  a symmetric ring structure also characteristic 
of a normal instanton, going through zero at the center of mass of the caloron.
The ring gets distorted as $\rho$ increases and dissociates for even larger 
$\rho$ into the two constituent BPS monopoles. This behaviour matches the 
one obtained analytically in  the $\rho \rightarrow 0 $ and $\rho 
\rightarrow \infty $ limits,  which are described in the next section.

\FIGURE{
\centerline{
\psfig{file=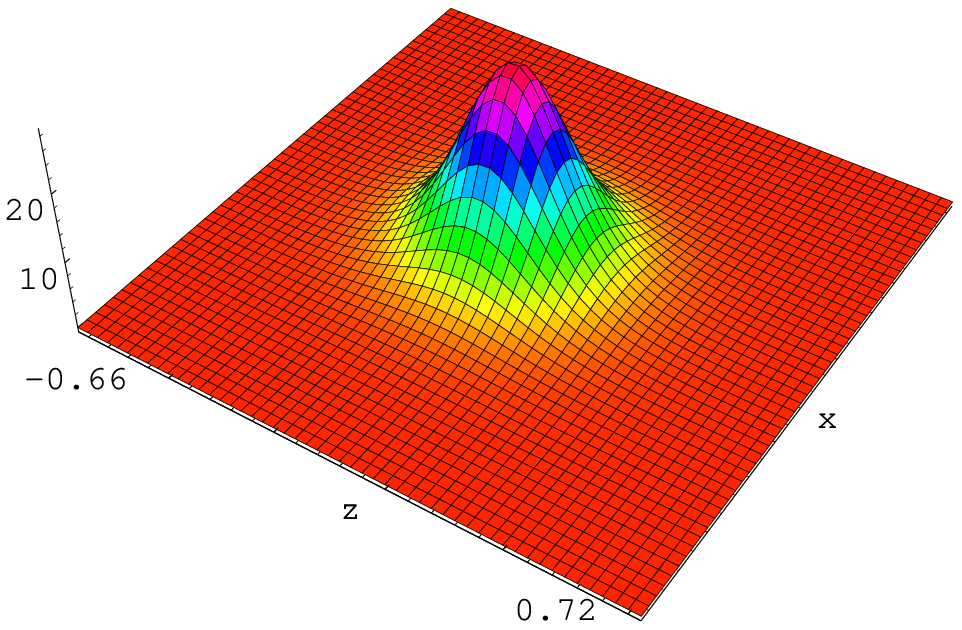,angle=0,width=8cm}
\psfig{file=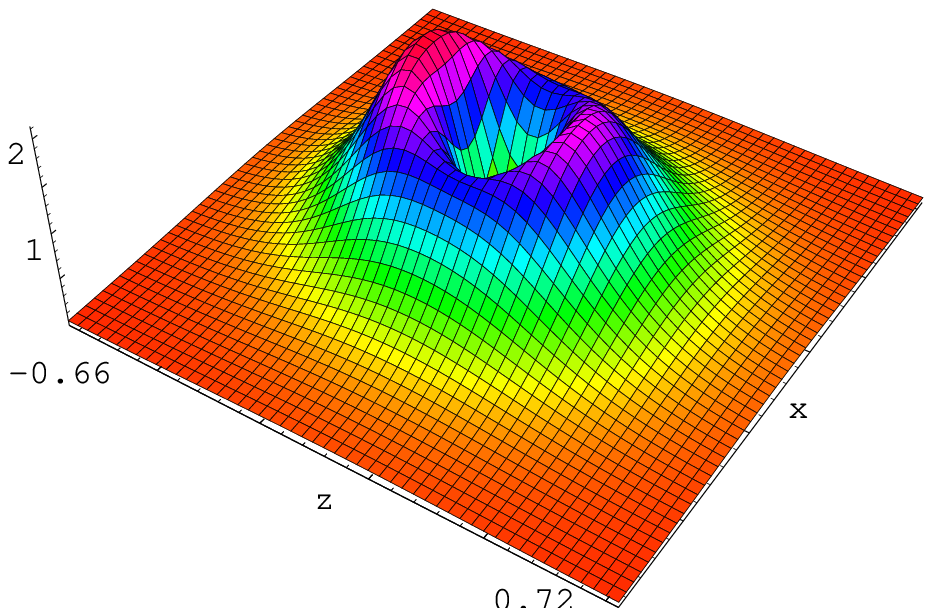,angle=0,width=8cm}}
\caption{The same as in Fig. \ref{fig1} but for $\rho=0.3$. }
\label{fig2}
}

\FIGURE{
\centerline{
\psfig{file=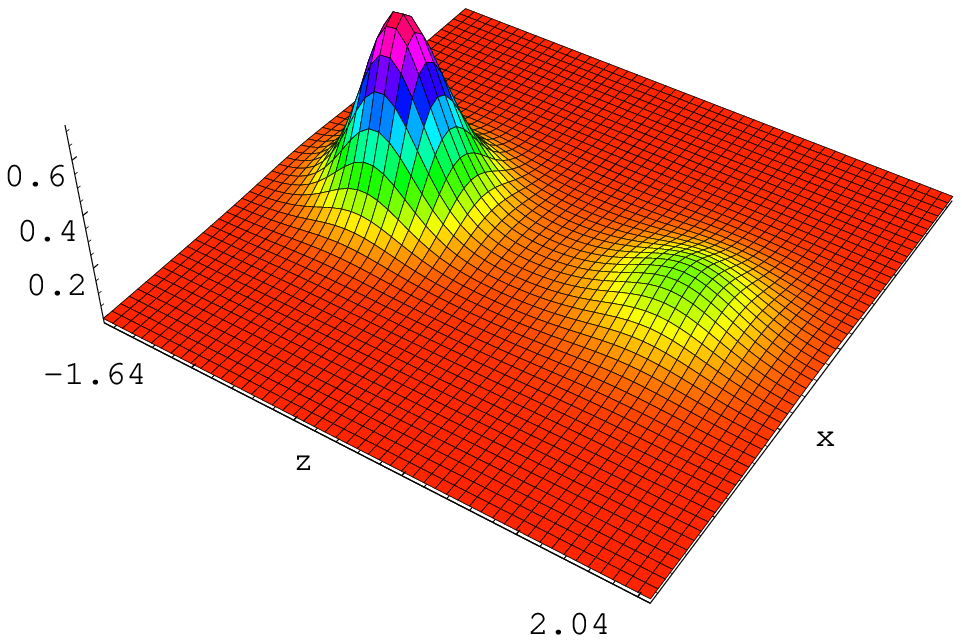,angle=0,width=8cm}
\psfig{file=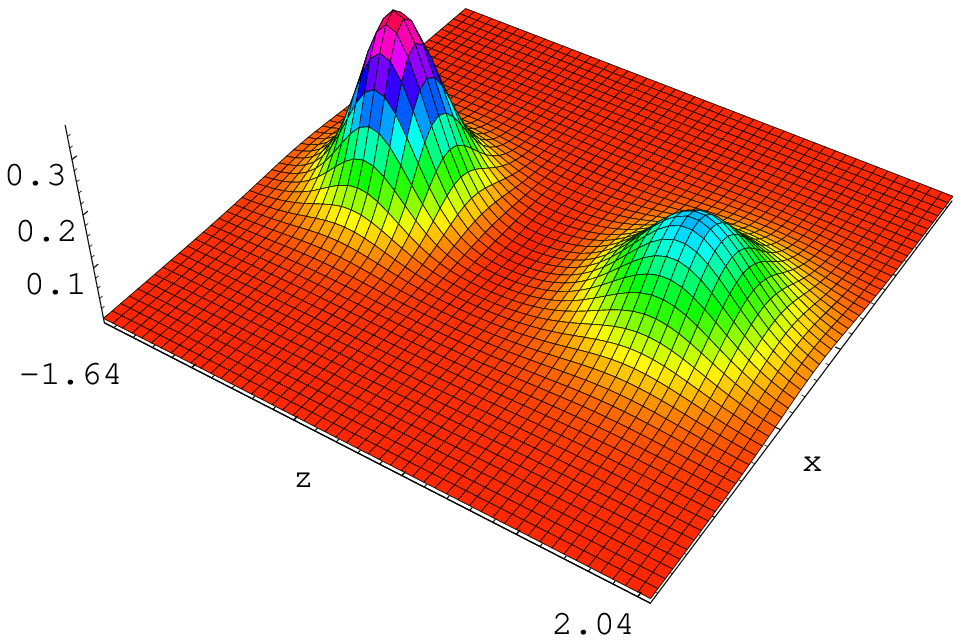,angle=0,width=8cm}}
\caption{The same as in Fig. \ref{fig1} but for $\rho=0.8$. }
\label{fig3}
}

\section{Properties of the solutions}
\label{s.prop}

In this section, we will  clarify the spatial structure of the modes by
studying their behaviour  in certain limits. Since calorons interpolate between
monopoles, instantons and HS calorons as one moves  along the moduli space, 
a similar phenomenon  is expected for the corresponding gluino zero-modes.

\subsection{$\rho \longrightarrow \infty$ limit}
There are several length scales involved in this problem ($m_a^{-1}$, 
the time period and the distance among constituents $\pi \rho^2$), 
so it is important  to clarify precisely in what regime we are thinking of. 
We will consider the simplest situation in which $\pi \rho^2$ is 
much larger than all other length scales in the problem. In addition, we 
will focus on the behaviour close to one of the monopoles $r_1<<\pi \rho^2$. 
Then,  for the other monopole one has $r_2\approx \pi \rho^2$. 

All our expressions are built in terms of $V$ and $U_a$ (see Eqs. \ref{vinv}
and \ref{usuba} in the appendix), so we will first 
analyse the behaviour of these quantities as a function of $\rho$. The
quantity $U_2$ behaves for large distances as $r_2/2\pi \longrightarrow
\rho^2/2$. Thus, $V$ becomes order $\rho^{-2}$. In order to compute the
leading behaviour of F in this limit one has to keep the first correction as
well. This is obtained by expanding 
\be
r_2=||\vec{r}_1-\pi \rho^2
\vec{k}||= \pi \rho^2 -(x_3-X^1_3) +\ldots
\ee
In this way we arrive at 
\be
V= \frac{1}{\rho^2} (1-\frac{1}{\rho^2}(U_1 -\frac{(x_3-X^1_3)}{2 \pi}))
\ee
and from here we see that $F$ is order $\rho^2$. From these results we
conclude that $W$ (Eq. \ref{wdoble}) is order $\rho^{-1}$, and $ \hat{\partial}W$ of order 
$\rho^{-3}$ while the derivatives of $U_a$ are order 1. Then  
it is easy to see that, of the different terms contributing to the zero-modes,
only the term  proportional to the $a=1$ BPS monopole field  is leading order. 
The latter is easily computable by realizing that the quantity inside 
parenthesis in Eq.~\ref{lambdaeq} now becomes precisely  $2/(F \rho^2)$. This leads 
to $\lambda_1=1$ and $\lambda_2=0$. The conclusion is that for
large separations and close to the center of the $a$th constituent monopole, 
the quantity $E_i^a$ is, after a performing a gauge rotation, 
simply given by the electric field of a single BPS monopole:
\be
E^{\rm BPS}(\vec{x}-\vec{X}^a; m_a)
\ee
Now we recall that the integral 
of the BPS monopole electric field square over space  equals $4 \pi m $. 
We reproduce, in this limit, the well-known fact that the contribution of 
each constituent monopole to the energy will then become, at sufficiently 
large separation,  proportional to $m_1$ and $m_2$ respectively.
Adding the two contributions to get the supersymmetric mode gives a
total integral of $8\pi^2$ as expected for the caloron.

Considering now the non-supersymmetric mode and taking into account that in 
Eq.~\ref{nonsusy} the first term is subleading in this limit, we conclude that
the contribution of the $a$th constituent monopole to the total energy
is given by $4 \pi \rho^4 m_1^2 m_2^2 /m_a$. This is consistent with the known
result~\cite{kvanbaal1} about the integrated densities which appear 
when computing the metric of the caloron moduli space.

\subsection{$\rho \longrightarrow 0$ limit}

To be precise the limit is defined by the requirement that 
$m_a r_a<<1$ and the distances $r_i$ are of order  $\rho$.
This implies that in the first approximation one can neglect 
the separation between the constituent monopoles ($\pi \rho^2 << r_i$).
As usual we will start by computing $U_a$ in this limit:
\be
U_a=\pmatrix{\frac{m_a x^2}{4 \pi} & \epsilon_a x_0 \cr 
\epsilon_a x_0 & \frac{1}{\pi m_a}}
\ee
where $x^2=x_0^2+r^2$ is the square of the 4d euclidean
distance to the origin.  
Notice that the 11 element of this matrix is negligible with respect 
to the 22 element, but has to be kept to be able to compute the inverse 
matrix. From here we conclude that 
to leading order $V=2 P_{11}/(x^2+\rho^2)$, where $P_{11}$ is the matrix 
(projector) whose only non-vanishing component is  11, equal to 1.
This projector appears in all our formulas and simplifies all vectors 
and matrices into single component. Using this it is easy to estimate
the contribution of both terms in the expression for 
$E_\mu^a$, Eq.~\ref{Eformula}. 
The second term turns out to be subleading with respect to the first one.
Using $\hat{\partial} (U_1+U_2)= \hat{x}$ 
the second term in this equation gives:
\be
\label{rhozero}
\frac{i m_a}{2 \pi} \frac{\rho^2}{(x^2+\rho^2)^2} \frac{\hat{x}
\bar{\sigma}_\mu \hat{x}}{x^2} + h.c. 
\ee
which is proportional ($m_a/(2 \pi)$) to the gauge field of an 
instanton in a certain  gauge. Hence, adding 
the $a=1$ and $a=2$ contributions, we conclude that the supersymmetric
zero mode reduces in this limit to the one of an BRST instanton.
Notice however that the combination of Eq.~\ref{rhozero} entering in the
non-supersymmetric zero mode vanishes.

For the case of the non-supersymmetric zero mode the leading term 
turns out to be the one proportional to the vector potential. For it we obtain
\be
\frac{A_\mu}{F}\longrightarrow -i \frac{\rho^2}{2 (x^2+\rho^2)^2}
\bar{\sigma}_\mu \hat{x}
+ \mbox{\rm h.c.}
\ee
Notice that this is precisely proportional to the
well-known non-supersymmetric  adjoint mode for the BRST instanton.

In summary,  in this limit both adjoint modes tend to the corresponding
ones for a BRST instanton.

\section{Conclusions}
\label{s.concl}

In this paper we have derived, following the ADHM formalism,  the analytic 
expressions for the gluino zero-modes 
of the Dirac operator in the background of topological charge 1 calorons
with non-trivial holonomy. These gluino zero-modes are relevant 
for semiclassical studies of 4D Super-symmetric Yang-Mills
theories. They can also turn out to be useful for analysing the 
structure and topological content of the QCD vacuum. In particular they 
can give a handle in the identification of constituent monopoles
inside instantons, a subject that has recently received much attention
\cite{gattringer1}-\cite{falkpierre}.

For $Q=1$ there are four linearly independent zero modes which come in pairs related 
by Euclidean CP transformations. Two of them correspond to the
super-symmetric zero modes and share the property that their density 
exactly reproduces the action density of the caloron. 
The other two modes show a quite distinct behaviour. For small scale parameter
$\rho$ they have the same ring structure as for trivial calorons or instantons
as exhibited in Fig. \ref{fig1}. As we increase $\rho$ 
the non-supersymmetric zero-mode rings get distorted 
(Fig. \ref{fig2}). For much larger $\rho$ both  zero mode
densities dissociate into two structures having the density profiles of 
BPS monopoles of masses $M_1$ and $M_2$ (Fig. \ref{fig3}).   
For the supersymmetric case the integrated contribution of each monopole to the energy
is proportional to its mass. However, for the non supersymmetric modes these 
contributions are inversely proportional to their masses.
By taking appropriate linear combinations it is possible to construct 
zero modes which, for intermediate to large scale parameters, single out 
only one of the two constituent monopoles.

All these features can certainly help in the identification
of constituent monopoles in lattice generated ensembles. 
This identification has been up to now performed solely on the basis 
of the action density and the properties of fundamental zero-modes
\cite{gattringer1,gattringer2}, \cite{ilgenfritz}, becoming particularly
complicated for low temperatures \cite{falkpierre}. 
Adjoint zero modes provide an additional tool for this analysis.
The supersymmetric zero mode density gives an estimate of the action 
density of the gauge field itself which is less sensitive to ultraviolet 
divergences.

Our present results can be extended to the case of SU(N) Q=1 calorons 
and to higher charge calorons. Furthermore, it is also possible to use
our techniques to construct gluino zero-modes which are anti-periodic 
in time, directly relevant for finite temperature ${\cal N}=1$ supersymmetry
\cite{falkus}.

\acknowledgments
We would like to thank Falk Bruckmann for discussions.
We acknowledge financial support from the Comunidad Aut\'onoma
de Madrid under the program PRICyT-CM-5 S-0505/ESP-0346.
M.G.P. acknowledges financial support from a Ram\'on y Cajal contract of
the MEC. M.G.P. is partially supported by the Spanish DGI under
contracts FPA2003-03801 and FPA2003-02877. A.G-A is partially supported 
by the Spanish DGI under contracts FPA2003-03801 and FPA2003-04597.

\section{Appendix}

In this appendix we will present the derivation the analytic 
formulas of the zero modes given in the text. 
All interesting quantities are expressed in terms  of the 
functions $u$ and $\OMF$ (see Eqs.~\ref{ueq}-\ref{OMeq}), 
and their integrals over z.
These functions belong  to a 4 dimensional vector  
space  over the field of quaternions. The space is 
the direct sum of the spaces  of functions annihilated (up to delta functions) 
by   the operator $\widetilde{M}\equiv\widetilde{A}-\hat{x}$ and its adjoint.
To obtain a basis of this space one must consider the functions
\bea
\Psi^{(+)}(z,x,\delta)=\chi(-\delta,\delta) \,
e^{i 2 \pi \bar{x}z}
\\
\Psi^{(-)}(z,x,\delta)=\chi(-\delta,\delta) \,
e^{i 2 \pi \hat{x}z}
\eea
They are to be taken as  periodic over $z$ ( and $x_0$) with unit period. The
function $\chi(-\delta,\delta)$ is the characteristic function
over the interval. 
In terms of these functions we can compute the 4 elements of the basis of 
the space: 
\be
\Psi^{\pm}_a \equiv \Psi^{(\pm)}(z-Z^a,x-X^a,\frac{m_a}{4 \pi}) 
\ee
The index  $a$  takes  two values (1 and 2), which can be thought as
labelling the two constituent monopoles. Indeed, $X^a$ stands for the location
and $m_a=4 \pi\delta_a$ for the mass (for $g^2=4 \pi$) of each of the monopoles.
The remaining 
coefficients are $Z^1=0$ and $Z^2=\frac{1}{2}$. 

To derive the formulas one needs to know how the operators $\widetilde{M}$
and $\widetilde{M}^\dagger$ act on them. These are given by 
\bea
\widetilde{M}^\dagger \Psi_a^{+} &=& \frac{-1}{2
\pi}\left( (e^a_1+ie^a_2)\delta(z-\delta_1)+
(e^a_1-ie^a_2)\delta(z+\delta_1)\right)\\
\widetilde{M} \Psi_a^{-} &=& \frac{-1}{2
\pi}\left( (\bar{e}^a_1+i \bar{e}^a_2)\delta(z-\delta_1)+
(\bar{e}^a_1-i\bar{e}^a_2)\delta(z+\delta_1)\right)\\
(\widetilde{M}-\widetilde{M}^\dagger) \Psi_a^{\pm} &=& 2 i
(\vec{x}-\vec{X}^a)\vec{\tau}\,  \Psi_a^{\pm} 
\eea
The quantities $e^a_1$ and $e^a_2$ are simple quaternionic 
functions of $x$ defined by the value of the basis functions at the
extreme of the interval:
\be
\lim_{z \rightarrow \delta_1^-} \Psi_a^{+}(z)=i\epsilon_a(e^a_1+ie^a_2)\equiv(\bar{e}^{\prime a}_1+i
\bar{e}^{\prime a}_2)
\ee
where $\epsilon_1=-1$ and $\epsilon_2=1$, and we have introduced new 
quaternionic quantities $e^{\prime a}_1$, $e^{\prime a}_2$. In all 
expressions a bar over a quaternion denotes its adjoint.

All of the necessary functions of $z$ entering the caloron formulas can be
expressed in terms of these functions. From the equations that they satisfy
(Eq. \ref{ueq}, \ref{OMeq}) we can easily deduce the general structure 
\bea
u&=&\sum_a u_a =\sum_a (\Psi_a^{+} A_a)\\
\OMF&=& \sum_a (\Psi_a^{+} D_a^{(+)} + \Psi_a^{-} D_a^{(-)}) 
\eea
where the  coefficients  ($A_a$, $D_a^{\pm)}$, etc) are  quaternionic 
functions of space-time. 

For all the necessary calculations of adjoint modes
one also needs the general form of $\hat{\partial}\OMF$:
\be
\hat{\partial}\OMF = \sum_a (4 \pi i (z-Z^a) \Psi_a^{+} D_a^{(+)} +\Psi_a^{+}
S_a^{(+)} + \Psi_a^{-} S_a^{(-)}) 
\ee
Finally, we need to know the integrals of the basis  functions over $z$. We
will need:
\be
I^{\pm}_\alpha = \int dz \Psi_b^{+ \dagger} \bar{\sigma}_\alpha \Psi_a^{\pm}=
\delta_{a b} \frac{m_a}{2 \pi} ({\cal P}^{ a\,  \alpha}_{\pm} \frac{1}{g(m_a r_a)} +
{\cal P}^{a \, \alpha}_{\mp})
\ee
and 
\be
\widetilde{I}_\alpha = 4 \pi i \int dz  z \Psi_b^{+ \dagger} \bar{\sigma}_\alpha
\Psi_a^{+}= -\frac{m_a^2}{2 \pi} \left(\frac{g'(m_ar_a)}{g^2(m_ar_a)}\right) {\cal P}^{ a\,
\alpha}_{+}(i \hat{n}_a)
\ee
where we have introduced  the quaternions
\be
{\cal P}^{ a\,  \alpha}_{\pm} =\frac{1}{2} (\bar{\sigma}^\alpha \pm
\hat{n}_a \bar{\sigma}^\alpha \hat{n}_a)
\ee
The symbol  $\hat{n}_a$ stands for a hermitian unitary traceless matrix 
defined through the decomposition
\be
(\vec{x}-\vec{X}^a)\vec{\tau}= r_a \hat{n}_a
\ee
where $r_a$ is the distance to the corresponding constituent monopole. 
The expressions also contain the function $g$:
\be
g(u)=u/\sinh(u)
\ee
and its derivatives evaluated at the product of the mass and the distance.

With the previous expressions one can compute the caloron vector potential
as well as the adjoint modes, once the coefficients $D^{(\pm)}_a$ and
$S_a^{(\pm)}$ are 
known. These can be deduced from the equations that define $u$ and $\OMF$ (Eqs.
\ref{ueq} and \ref{OMeq}), but now the whole analytic structure reduces to a finite-dimensional 
linear problem in quaternions.  Essentially, this follows from the matching 
at the edges of the intervals $z=\pm \delta_1$. For example, the absence of 
derivatives of delta functions in the equation for $\OMF$, implies that this
function must me continuous at $z=\pm \delta_1$. If we note
$\OMF(\delta_1)=W_1+iW_2$, then we can 
compute the coefficients of $u$ and $\OMF$ in terms of $W_i$ by the
continuity equations:
\bea
W&=&\bar{e}^{\prime a} D_a^{(+)} + e^{\prime a} D_a^{(-)} \\
\hat{\partial}W&=& -4 \pi \delta_a e^a D_a^{(+)}+ \bar{e}^{\prime a} S_a^{(+)} + e^{\prime a} S_a^{(-)}
\eea
The equation has been rewritten as a vector equation in terms of two
(quaternionic)  component column vectors $W$, $e^{\prime a}$, $\ldots$ . 
Notice that each equation is valid for both values of $a$. With ordinary
vector space techniques one can solve for the coefficients:
\bea
D_a^{(+)}&=& \frac{ i \hat{n}_a g(m_a r_a)}{m_a r_a} e^{a \dagger} W\\
D_a^{(-)}&=& \frac{ - i \hat{n}_a g(m_a r_a)}{m_a r_a} \bar{e}^{a \dagger} W\\
S_a^{(+)}&=& \frac{ i \hat{n}_a g(m_a r_a)}{m_a r_a} e^{a \dagger}
\hat{\partial}W - \frac{ m_a \cosh(m_a r_a)}{\sinh^2(m_a r_a)} e^{a \dagger}
W\\
S_a^{(-)}&=& \frac{ -i \hat{n}_a g(m_a r_a)}{m_a r_a} \bar{e}^{a \dagger}
\hat{\partial}W + \frac{ m_a }{\sinh^2(m_a r_a)} e^{a \dagger} W
\eea
The  coefficient $A_a$ appearing in the expansion of $u$ can be related 
to $D_a^{(+)}$ by the equation $\widetilde{M} \OMF =u/F$. Hence, we get
\be
A_a= 2 i r_a F \hat{n}_a D_a^{(+)}
\ee

Combining all the previous formulas  we arrive at  
\be
\label{maineq}
u_a^\dagger \bar{\sigma}_\alpha \hat{\partial}\OMF_a = W^\dagger {\cal L}_{a
\alpha} W +  W^\dagger \widetilde{{\cal L}}_{a} \bar{\sigma}_\alpha \hat{\partial}W 
\ee
where 
\be
\label{Leq}
{\cal L}_{a \alpha} = \frac{2 F m_a g(m_a r_a)}{\pi } e^a \left( -i \frac{g^2(m_a
r_a)-1}{2 m_a^2 r_a^2}  {\cal P}^{ a\,  \alpha}_{+} -i 
\frac{g'(m_a r_a)}{2 m_a r_a}  {\cal P}^{ a\,  \alpha}_{-}  \right) e_a^\dagger 
\ee
and 
\be
\label{LPeq}
\widetilde{{\cal L}}_{a }=\frac{-iF g(m_a r_a) }{\pi m_a r_a} e_a
\hat{n}_a \left( - g(m_a r_a) \bar{e}_a^\dagger  + e_a^\dagger  \right)
\ee

Up to this point all expressions seem to depend only on a single distance
$r_a$. The mixing among the two coordinates and the relation between the 
constituent monopoles is hidden in the expression of
$W$. The main equation satisfied by $W$ is 
\be
\frac{F}{\pi} \sum_a \frac{g(m_a r_a)}{m_a} e_a e_a^\dagger W =
\frac{\rho}{2} \pmatrix{1 \cr -i \tau_3}
\ee
From here one can solve for $W$. 
It is very easy to realise that $W$ must be
a linear combination of the quaternion $i\tau_3$ and unity. This follows from 
the equation  $\OMF=R^{-1} q$. Since $q$ is a combination of these two
quaternions and $R$ commutes with quaternions (and is therefore real), this
property extends to $W$. In summary, we have 
\be
W=\frac{\rho}{2} V \pmatrix{1 \cr -i \tau_3} 
\label{wdoble}
\ee
where V is a real $2 \times 2$ matrix, whose inverse is sum of contributions
from 
the two constituent monopoles
\be
V^{-1}= \frac{\rho^2}{2} + \sum_a U_a
\label{vinv}
\ee
with
\be
U_a=\frac{g(m_a r_a)}{2 \pi m_a} \pmatrix{\cosh(m_a r_a)-\cos(m_a x_0) &
\epsilon_a \sin(m_a x_0) \cr \epsilon_a \sin(m_a x_0) & \cosh(m_a
r_a)+\cos(m_a x_0)} 
\label{usuba}
\ee
An interesting relation between $U$ and the vectors $e^a$ is given by
\be
 U_a = \frac{1}{\pi} \frac{g(m_a r_a)}{m_a} e_a e_a^\dagger
-\frac{\epsilon_a r_a}{2 \pi} \pmatrix{0 & i\hat{n}_a \cr  -i\hat{n}_a   & 0}
\ee

Now we have all the ingredients to calculate all the relevant quantities
concerning calorons, including the adjoint zero-modes. For example, we can 
obtain the scalar function $F$ 
\be
F= \frac{1}{1-\rho^2\mbox{ \rm Tr}(V)/2}
\ee
For the vector potential, we can write the following expression:
\be
A_\mu=
\frac{-i F}{2} W^\dagger  \bar{\sigma}_\mu
\hat{\partial} (U_1+U_2)\,  W
+\mbox{ \rm h. c. } 
\ee

Finally, we proceed to the computation of the basic quantity which enters 
into the expression of the zero modes $E^a_\alpha(x)$. This is obtained 
from  Eq.~\ref{maineq} after multiplying by $i$ and taking the hermitian
part. The result is the sum of two terms. The first one has a very transparent
interpretation. To show this, one must first realize that the traceless part
(or hermitian part) of the quantity inside parenthesis in Eq.~\ref{Leq} is precisely
$ E^{\rm  BPS}_\alpha(x-X^a; m_a)/m_a^2$, where $ E^{\rm  BPS}_\alpha(x-X^a; m_a)$ is the gauge field of a BPS monopole of mass $M_a$
centered at one of the constituent monopoles. This is 
sandwiched between  the quaternion $Q_a\equiv e^{a \dagger} W$ and its adjoint. 
If we write 
\be
Q_a\equiv e^{a \dagger} W= |Q_a| \Omega^\dagger_a
\ee
where $\Omega_a$ is a unitary matrix and we define
\be
\lambda_a= |Q_a|^2 \frac{2 F g_a}{\pi m_a}
\ee
then we conclude that the first term in $E^a_\alpha(x)$ is given by 
\be
\lambda_a  \Omega_a E^{\rm  BPS}_\alpha(x-X^a; m_a) \Omega_a^\dagger
\ee
which is just the field of a BPS monopole, gauge rotated and weighted by 
$\lambda_a$. The weight factors are positive and satisfy:
\be
\sum_a \lambda_a= (1-\frac{1}{F})
\ee
An explicit formula to compute them in terms of the matrix $V$ is
\be
\label{lambdaeq}
\lambda_a= \frac{F\rho^2}{2}\left(\mbox{\rm Tr}(VU_aV) +\frac{\epsilon_a}{2
\pi} (x_3-X_3^a)(\mbox{\rm Tr}^2(V)-\mbox{\rm Tr}(V^2))\right) 
\ee

The second piece contributing to  Eq.~\ref{maineq} is also simplified 
if one realizes that 
\be
 \widetilde{{\cal L}}_{a }=-F \bar{\partial} U_a
\ee
Hence, one reaches to the following simple formula for the quantity 
$E^a_\mu$, representing the contribution of constituent monopole $a$ to the field
strength
\be
\label{Eformula}
E^a_\mu  = \frac{F g(m_a r_a)}{\pi m_a} W^\dagger e^a E^{\rm 
BPS}_\mu(x-X^a; m_a)\,  e^{a \dagger} W -\frac{i F}{2} W^\dagger \bar{\partial} U_a \bar{\sigma_\mu}
\hat{\partial} W + \mbox{\rm h.c.}
\ee
The time-like component ($\mu=0$) of the first term vanishes. For the second
term we have $E^1_0=-E^2_0$. This is easily concluded by realizing 
that 
\be
\hat{\partial} W= -V \hat{\partial} (U_1+U_2) W
\ee

\end{document}